\documentclass[%
 reprint,
superscriptaddress,
showkeys,
 amsmath,amssymb,
 aps,
prx,
floatfix,
]{revtex4-2}

\usepackage{xcolor}
\definecolor{webblue}{HTML}{000099}
\usepackage[colorlinks=true, linkcolor=webblue, urlcolor=webblue, citecolor=webblue]{hyperref}
\usepackage{enumitem}
\usepackage{clipboard}
\usepackage{changes}
\usepackage{graphicx}
\usepackage{dcolumn}
\usepackage{bm}
\usepackage{siunitx}
\usepackage{xcolor}
\usepackage{multirow}

\begin{document}

\title{Feasibility study of continuous electronic Pomeranchuk cooling with a flavor-degenerate Wigner crystal}

\author{Robin J. Dolleman}
 \email{dolleman@physik.rwth-aachen.de}
 \affiliation{JARA-FIT and 2nd Institute of Physics, RWTH Aachen University, 52074 Aachen, Germany}
 \affiliation{HFML-FELIX, Toernooiveld 7, 6525ED Nijmegen, The Netherlands}
 \affiliation{Radboud University, Institute for Molecules and Materials, Heyendaalseweg 135, 6525 AJ, Nijmegen, The Netherlands}
 \author{Ammon Fischer}
\affiliation{Institute for Theory of Statistical Physics, RWTH Aachen University, and JARA Fundamentals of Future Information Technology, 52062 Aachen, Germany}
\affiliation{Max Planck Institute for the Structure and Dynamics of Matter, Center for Free Electron Laser Science, 22761 Hamburg, Germany}
\author{Lennart Klebl}
\affiliation{I. Institute for Theoretical Physics, University of Hamburg, Notkestraße 9-11, 22607 Hamburg, Germany}
\affiliation{Institute for Theoretical Physics and Astrophysics, University of Würzburg, Am Hubland, 97074 Würzburg, Germany}
\author{Alexander Rothstein}
\affiliation{JARA-FIT and 2nd Institute of Physics, RWTH Aachen University, 52074 Aachen, Germany}%
\affiliation{Peter Gr\"unberg Institute (PGI-9), Forschungszentrum J\"ulich, 52425 J\"ulich, Germany}
\author{Dante M. Kennes}
\affiliation{Institute for Theory of Statistical Physics, RWTH Aachen University, and JARA Fundamentals of Future Information Technology, 52062 Aachen, Germany}
\affiliation{Max Planck Institute for the Structure and Dynamics of Matter, Center for Free Electron Laser Science, 22761 Hamburg, Germany}
\author{Bernd Beschoten}
\affiliation{JARA-FIT and 2nd Institute of Physics, RWTH Aachen University, 52074 Aachen, Germany}
\author{Florian Libisch}
\affiliation{Institute for Theoretical Physics, TU Wien, 1040 Wien, Austria}
\author{Christoph Stampfer}
 \email{stampfer@physik.rwth-aachen.de}
\affiliation{JARA-FIT and 2nd Institute of Physics, RWTH Aachen University, 52074 Aachen, Germany}
\affiliation{Peter Gr\"unberg Institute (PGI-9), Forschungszentrum J\"ulich, 52425 J\"ulich, Germany}

\begin{abstract}
Achieving sub-millikelvin electron temperatures in nanoelectronic devices could unveil new transport phenomena, extend quantum coherence times, and enhance the precision of quantum metrology.
However, maintaining such low temperatures continuously remains a long-standing challenge.
Here, we propose and simulate an on-chip cooling cycle that harnesses the entropy difference between an electron liquid (EL) and a Wigner crystal (WC) in flavor-degenerate flat-band materials.
Cooling is driven by a current through a device with a locally gated region.
Within this region, the charge carrier density is tuned such that a WC forms beneath the gate.
As carriers transition from an EL to WC phase, their entropy increases, extracting heat and the sliding WC advects this heat along the device.
The heat is then released when carriers transition back to the EL phase, which establishes distinct hot and cold regions and a steady temperature gradient over the device.
Simulations show net cooling for sufficiently low current densities typically below 1~\si{\nano\ampere\per\micro\meter}, whereas Joule heating dominates at higher currents.
Within the gated region, we estimate cooling powers of up to \SI{8.4}{\atto\watt\per\micro\meter} at a bath temperature of \SI{4}{\milli\kelvin}.
Our approach can achieve electron temperatures well below \SI{1}{\milli\kelvin} under suitable conditions, promising a route towards continuous on-chip cooling in this temperature regime.
Our approach applies to any flat-band material with low-energy flavor degeneracy (valley and/or orbital) and low disorder, including gapped Bernal-stacked bilayer graphene, rhombohedral-stacked multilayer graphene, and magic-angle twisted bilayer graphene.
\end{abstract}

\maketitle

\section{Introduction}
Electron temperatures below \SI{1}{\milli\kelvin} in micro- and nanoelectronic devices open the door to probe exotic quantum phases that occur at very small energy scales~\cite{Jones2020Dec}.
 These include, for example, fractional (anomalous) quantum Hall states~\cite{Chesi2008Oct,Pan2015Jan,Samkharadze2015Feb,Werkmeister2024Mar,Park2023Oct,Lu2024Feb}, non-Abelian states of matter~\cite{Nayak2008Sep,Stern2010Mar}, ordered spin states~\cite{Simon2007Apr,Simon2008Jan,Braunecker2013Oct,Scheller2014Feb}, the ground state of Wigner crystals (WCs) at zero magnetic field~\cite{Yoon1999Feb,Knighton2018Feb,Falson2022Mar,Hossain2020Dec,Spivak2004Oct,Spivak2003Mar,Spivak2006Sep}, and certain superconducting phases~\cite{Schuberth2016Jan,Zhou2022Jan,Mackenzie2003May}. 
Lower temperatures also help improve thermalization, which is key to reducing errors in semiconductor and superconducting quantum devices~\cite{Makhlin2001May,Hanson2007Oct,Clarke2008Jun,Pekola2013Oct}, and enhancing the precision of quantum metrology tools like quantum Hall resistance standards~\cite{Poirier2009Jun,Ribeiro-Palau2015Nov} and charge pumps~\cite{Piquemal2000Jun,Giblin2012Jul,Connolly2013Jun}.
\begin{figure*}
    \centering \includegraphics{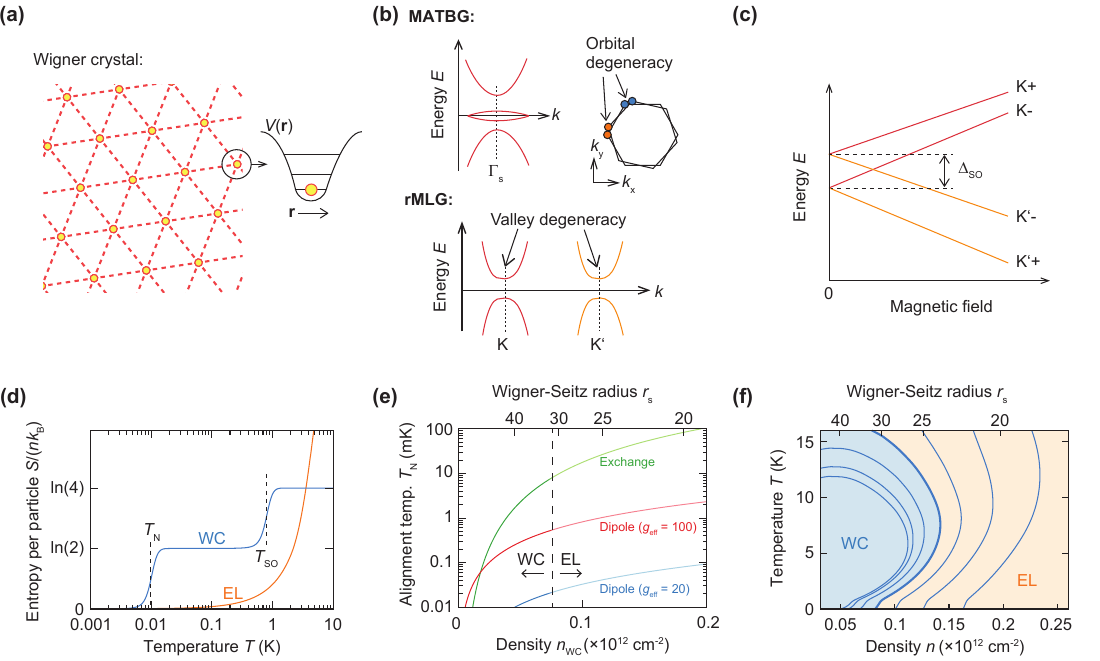}
    \caption{(a) Illustration of the triangular lattice of charge carriers in the WC. Each charge carrier in confined within a potential created by the surrounding carriers. (b) Exemplary band structures of magic-angle twisted bilayer graphene (MATBG) and gapped rhombohedrally stacked multilayer graphene (rMLG); for two layers, rMLG corresponds to Bernal-stacked bilayer graphene. Depending on the system, each carrier can carry an orbital and/or valley degree of freedom close to the band gaps. (c) Evolution of the four lowest energy states of a confined charge carrier in a valley-degenerate system as a function of out-of-plane magnetic field (schematically).  $K$ and $K'$ indicate the valleys and $+$/$-$ indicates the spin direction. (d) Entropy per particle versus temperature for both the Wigner crystal (WC) phase and electron liquid (EL) for $n = \num{9.75e10}~$cm$^{-2}$ in a graphene-like flat band system, assuming $m^*/m_{\rm e} = 0.4$. (e) Estimated alignment temperature $T_{\mathrm{N}}$ within a Wigner crystal versus charge carrier density $n_{\mathrm{WC}}$ for the case of exchange and dipole interactions with different values of $g_\mathrm{eff}$. The vertical dashed line corresponds to $r_\mathrm{s} = 31$. (f) Coexistence lines between the WC and EL phases in the density-temperature phase diagram (blue lines) for a graphene-like system near a band edge and assuming an effective mass $m^*/m_{\rm e} = 0.4$. Only lines with $r_s < 50$ are included due to the applicability of the model for the chemical potentials \cite{Drummond2009Mar}.}
    \label{fig:pomeranchuk_nu4}
\end{figure*}

At present, most nanoelectronic devices are cooled using dilution refrigerators, which rely on the phase separation and entropy difference in a liquid $^3$He/$^4$He mixture~\cite{Zu2022Jan}. 
However, since electronic devices are typically electrically isolated from the refrigerator, cooling must be achieved via an electrically insulating substrate.
In this indirect pathway, heat flows only via electron--phonon and phonon--phonon processes, whose thermal conductances drop rapidly on approaching sub-mK temperatures~\cite{Casparis2012Aug,McKitterick2016Feb,Karvonen2004Nov,Wellstood1994Mar,Giazotto2006Mar}.
As a result, the lowest continuously achievable temperature in nanoelectronic devices with dilution refrigeration remains about \SI{4}{\milli\kelvin}~\cite{Bradley2016Jan,Xia2000May,Samkharadze2011May,Iftikhar2016Sep}, and reaching sub-mK electron temperatures is challenging. 

To overcome this limitation, researchers adopt on-chip refrigeration methods that cool the electrons directly, circumventing the unfavorable thermal resistances inherent to cooling through the substrate~\cite{Jones2020Dec,Muhonen2012Mar}.  
Several approaches rely on quantum dots or superconducting tunnel junctions as energy filters to selectively remove hot electrons~\cite{Edwards1993Sep,Giazotto2006Mar,Rajauria2007Jul,Prance2009Apr,Muhonen2012Mar,Vischi2019Mar,Rabani2008Jul}.
However, extending these techniques to the sub-millikelvin regime faces significant challenges.
For superconducting tunnel junctions, the cooling efficiency is limited by non-equilibrium quasiparticle fluctuations~\cite{deVisser2011Apr}, subgap leakage in the superconducting density of states, and slow electron–electron relaxation~\cite{Pekola2004Feb,Prance2009Apr}.
Quantum-dot–based refrigerators, in contrast, require extremely stable and narrow energy levels and are therefore highly sensitive to charge noise originating from the host material or transmitted through the measurement circuitry~\cite{Camenzind2018Aug,Maradan2014Jun}.
An alternative on-chip approach exploits adiabatic demagnetization of a metallic refrigerant, cooling the sample leads via hyperfine coupling between electrons and nuclear spins~\cite{Palma2017Dec,Bradley2017Apr,Autti2023Aug,Yurttagul2019Jul,Sarsby2020Mar}.
While combining on-chip and off-chip adiabatic demagnetization has enabled sub-mK electron temperatures~\cite{Sarsby2020Mar}, this technique is inherently one-shot and sustains ultralow temperatures only for a limited duration (up to 85 hours).

Theoretically, quantum-confinement–based cooling schemes have been proposed, in which a gate-controlled adiabatic redistribution of electrons into additional sub-bands decreases the electron temperature ~\cite{Autti2025Apr}.
However, the implementation remains single-shot and continuous sub-mK cooling would require a considerable scaling of the device architecture.
More recently, an entropy-based cooling mechanism has been proposed, in which electrons are driven through a high-entropy region and thereby extract heat from an electronic bath as their entropy increases~\cite{Aliani2025Nov}.
This concept relies on the enhanced electronic entropy associated with nodal structures in $\pi$-phase-biased superconducting tunnel junctions.
Experimentally realizing this mechanism remains nontrivial, as it requires sufficiently precise control of the phase bias and junction parameters, while limiting disorder broadening to preserve the entropy-enhancing density-of-states features.

In this work, we propose a continuous on-chip refrigeration technique that similarly exploits large electronic entropy, but instead derives it from a phase transition between a two-dimensional (2D) electron liquid (EL) and a flavor-degenerate Wigner crystal (WC).
A WC is a phase of matter emerging when charge carriers crystallize into a lattice due to Coulomb repulsion dominating over kinetic energy at sufficiently low carrier densities and temperatures [Fig.~\ref{fig:pomeranchuk_nu4}(a)]~\cite{Grimes1979Mar,Wigner1934Dec}. 
WCs are typically realized in very clean 2D systems and can be further stabilized by flat electronic bands, which may arise intrinsically or be generated by Landau level formation in strong magnetic fields~\cite{Knighton2018Feb,Hossain2020Dec,Spivak2004Oct,Spivak2003Mar,Spivak2006Sep,Andrei1988Jun,Yoon1999Feb,Monarkha2012Dec,Smolenski2021Jul,Falson2022Mar,Tsui2024Apr,Seiler2025Oct}.
At zero magnetic field, the WC features a large entropy because the exchange coupling $J$ between localized carriers is exponentially suppressed as their separation increases, leaving nearly degenerate local moments at each lattice site~\cite{Roger1984Dec,Bernu2001Jan}.
As a result, heating or a magnetic field stabilises the WC phase via a mechanism analogous to the Pomeranchuk effect in $^3$He~\cite{Spivak2003Mar,Spivak2004Oct,Spivak2010May}.
We demonstrate here that these properties allow the Wigner crystal to extract heat and thereby reduce the electron temperature of an electronic device.

In our approach, we consider a 2D, flavor-degenerate material with a hexagonal lattice, representative of graphene-like systems. 
Here, "flavor" denotes the low-energy internal degeneracy: near the K/K' points of rhombohedrally stacked graphene this is commonly valley~\cite{Koshino2010Mar}, whereas near the moir\'e-induced band gap in magic-angle twisted bilayer graphene (MATBG) near the center of the mini-Brillouin zone this degeneracy is orbital [Fig.~\ref{fig:pomeranchuk_nu4}(b)]~\cite{Andrei2020Dec}. 
To obtain typical parameter values for a flat-band material, such as the electron-liquid entropy and the effective charge-carrier mass, we use experimental estimates from MATBG near full filling of the moiré lattice, assuming a moiré superlattice density $n_\mathrm{s} = \num{2.6e12}\,\mathrm{cm}^{-2}$, which corresponds to a twist angle $\theta \approx 1.1^{\circ}$~\cite{Rozen2021Apr,Cao2018Apr,Cao2018Apr2}.

We first discuss the WC entropy in multilayer graphene and, using the Clausius-Clapeyron equation, derive the phase diagram versus charge carrier density~$n$ and temperature~$T$ (Sec.~\ref{sec:phasediagram}).  
Next, we propose a thermodynamic cooling cycle that exploits the phase transition between the EL and the WC (Sec.~\ref{sec:cycle}).
Finite element simulations indicate that this cooling cycle could be highly effective at ultralow temperatures, cooling electrons to below 1~mK while achieving cooling powers of up to $\sim$8~\si{\atto\watt\per\micro\meter}.
A successful implementation of this device scheme would constitute a significant advance in the continuous cooling of electron temperatures, opening new opportunities in quantum computing, electronic transport studies, as well as quantum sensing and metrology.

\section{Entropy of a Wigner crystal and electron liquid}\label{sec:phasediagram}
The strength of interactions between itinerant charge carriers is characterized by the dimensionless Wigner-Seitz radius $r_s = a/a_{\mathrm{B}}$~\cite{Tanatar1989Mar}, where $a_B$ is the (effective) Bohr radius,
\begin{equation}\label{eq:WignerSeitz}
    a_{\mathrm{B}} = \frac{4 \pi \varepsilon_0 \varepsilon_r' \hbar^2}{m^* e^2},
\end{equation}
and $a = (\pi n)^{-1/2}$ is the average distance between charge carriers. 
Here, $n$ is the charge carrier density, $m^*$ is the charge carrier effective mass, $e$ the elementary charge, $\varepsilon_0$ the vacuum permittivity, $\varepsilon_r'$ the relative in-plane dielectric constant, and $\hbar$ the reduced Planck constant. 
Quantum Monte Carlo simulations indicate that at $T = 0$ the EL becomes unstable and a triangular WC forms for $r_s \gtrsim 31$~\cite{Drummond2009Mar}.

In the WC phase, each carrier occupies the ground state of the electrostatic potential well formed by its neighbors~[see Fig.~\ref{fig:pomeranchuk_nu4}(a)].
As shown in Fig.~\ref{fig:pomeranchuk_nu4}(b), depending on the system the ground state can be orbital and/or valley degenerate. 
Spin–orbit coupling lifts the spin degeneracy, resulting in a flavor-degenerate ground state. 
As a concrete and well-studied example we consider gapped Bernal bilayer graphene, where the spin-orbit gap $\Delta_{\rm SO}$ typically has values in the range \SIrange{40}{80}{\micro\electronvolt}~\cite{Banszerus2020May,Banszerus2021Sep,Kurzmann2021Oct,Tong2022Feb,Banszerus2025Jul}.
This corresponds roughly to a temperature scale of $T_\mathrm{SO} \approx \SIrange{0.5}{0.9}{\kelvin}$. 
In the absence of interactions between neighboring carriers, the flavor degeneracy holds at zero magnetic field~\cite{Koshino2010Mar}. 

When neighboring carriers interact, however, this degeneracy can be lifted. 
In a WC, this can occur through exchange processes arising from the Pauli exclusion principle, as well as through magnetic dipole–dipole interactions between the carriers.
If exchange (ex) dominates, the relevant energy scale is set by the exchange coupling, which scales as $J \propto \exp(-\gamma \sqrt{r_s})$ with $\gamma \approx 1$~\cite{Spivak2004Oct,Roger1984Dec}, and defines the flavor alignment temperature $T_\mathrm{N,ex}$ via $k_\mathrm{B}T_\mathrm{N,ex} = J$. 
When $k_\mathrm{B}T > J$, the local flavors are effectively free.
Figure~\ref{fig:pomeranchuk_nu4}(e) estimates $T_\mathrm{N,ex}$ versus $n$, assuming pair exchange with energy scale $J_2$ dominates and $J_2 = k_\mathrm{B} T_\mathrm{N,ex}$ (see green trace)~\cite{Hirashima2001May}. 
We note that on a triangular WC, three-particle ring exchange $J_3$ can compete with $J_2$ due to geometric frustration~\cite{Bernu2001Jan}, but their magnitudes are similar, so we take $J_2$ as a representative scale. 
As $n_\mathrm{WC}$ increases, the average distance $a$ and $r_s$ decrease, enhancing exchange and raising $T_\mathrm{N,ex}$. 
Typically, $T_\mathrm{N,ex}$ is of order 1--10~mK for $a$ on the order of tens of nanometres, while in a more dilute WC it can fall well below 1~mK.

If instead dipole–dipole (dd) interactions dominate, the alignment temperature is set by the characteristic scale
\begin{equation}
T_{\mathrm{N,dd}} \sim \frac{z\mu_0}{4\pi k_\mathrm{B}} \frac{(g_\mathrm{eff} \mu_\mathrm{B})^2}{4a^3}.
\end{equation}
Considering an effective $g$-factor in the range of $g_\mathrm{eff} = 20$--$100$, based on experimental results on gapped Bernal stacked bilayer graphene~\cite{Tong2021Jan,Moller2023Sep}, we find that dipole interactions only become important at low $n_\mathrm{WC}$ and large orbital magnetic moments [see red and blue traces in Fig.~\ref{fig:pomeranchuk_nu4}(e)].
Consequently, we expect that in all practical devices, the dipolar interactions are subleading and in the following we assume that exchange driven flavor ordering determines the alignment temperature ($T_\mathrm{N,ex} = T_\mathrm{N}$). 

In addition to the flavor degeneracy, the WC can also carry entropy through lattice vibrations (phonons).
We investigate this contribution in Appendix C and find that at temperatures $\lesssim 10$~K the flavor contribution dominates, while phonon contributions remain small.
Thus, we expect that for $T_\mathrm{N} < T < T_\mathrm{SO}$, the specific entropy is $S_\mathrm{WC} = n k_\mathrm{B} \mathrm{ln}(2)$ due to valley or orbital degeneracy, while for $T > T_\mathrm{SO}$ it increases to $S_\mathrm{WC} = n k_\mathrm{B} \mathrm{ln}(4)$. 
Figure~\ref{fig:pomeranchuk_nu4}(d) illustrates the expected entropy per particle as a function of temperature for the WC phase, where we ignore the small phonon contribution to the entropy for simplicity.
Importantly, for a sufficiently low $n_\mathrm{WC}$ the WC phase retains high entropy even at sub-mK temperatures, a key factor that enables the continuous cooling cycle proposed in Sec.~\ref{sec:cycle}.

\subsection{Pomeranchuk effect}
In this subsection, we examine how entropy shapes coexistence between the WC and EL phases near the melting transition. 
In contrast to the WC, the EL has suppressed low-$T$ entropy because interactions lift the flavor degeneracy of the itinerant carriers.
To represent the EL entropy, we use the entropy data calculated for MATBG within the mean-field Dirac-revival model of Ref.~\cite{Rozen2021Apr}, interpolating these data as a function of $n$ and $T$. 
The resulting EL entropy is substantially smaller than that of the WC at low $T$, as illustrated in Fig.~\ref{fig:pomeranchuk_nu4}(c) for $n=\num{9.75e10}~\mathrm{cm}^{-2}$.
This entropy difference controls the shape of the phase diagram.

As noted in Ref.~\cite{Spivak2006Sep}, a sharp first- or second-order EL--WC transition is precluded; instead an intermediate microemulsion phase appears.
Moreover, small quantum corrections render the exact value of $r_s$ where melting occurs not well constrained. 
Accordingly, we approximate the EL--WC transition by coexistence lines in the $n$ versus $T$ phase diagram.
Along these lines, chemical potentials and volumetric fractions of the two phases remain constant, and their slopes are determined by the Clausius-Clapeyron relation (derived in the Appendix A): \begin{equation}\label{eq:ClausiusClapeyron}
 \frac{\mathrm{d}n}{\mathrm{d}T} = - \frac{S_\mathrm{WC} - S_\mathrm{EL}}{\mu_\mathrm{WC} - \mu_\mathrm{EL}} ,
\end{equation}
where $S_\mathrm{WC}$ and $S_\mathrm{EL}$ are the entropy densities of the WC and EL phase, and $\mu_\mathrm{WC}$ and $\mu_\mathrm{EL}$ their chemical potentials. 
We plot the coexistence lines by numerically integrating Eq.~\eqref{eq:ClausiusClapeyron} in the $n$ versus $T$ phase diagram, starting from specific values of $r_s$ at $T=0$~K, as shown in Fig.~\ref{fig:pomeranchuk_nu4}(d).
The thick blue line represents the coexistence line for $r_s = 31$, with additional lines spaced at $\Delta r_s = 3$ to depict the broader region of phase transition.
Because $\Delta S = S_\mathrm{WC} - S_\mathrm{EL}$ changes sign with increasing $T$, the slope ${\mathrm{d}n}/{\mathrm{d}T}$ also changes sign.
Thus, at certain fixed densities $n$ the WC fraction increases with temperature, essentially ``freezing by heating", underscoring the central role of the entropy difference.
This behavior, the electronic analogue of the Pomeranchuk effect in $^3$He~\cite{Richardson1997Jul}, has been predicted when a WC competes with an EL under weak spin–orbit coupling~\cite{Spivak2004Oct,Spivak2010May}.

\section{Cooling cycle}\label{sec:cycle}
\begin{figure*}
    \centering \includegraphics{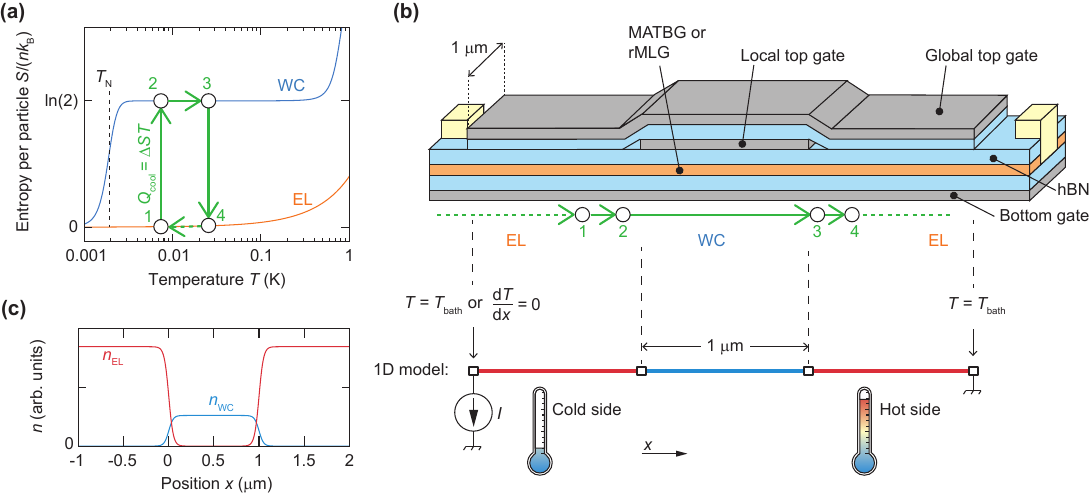}
    \caption{(a)~Sketch of the proposed thermodynamic cycle as an ideal isothermal process in the entropy–temperature diagram. (b)~Schematic of the locally top-gated multilayer graphene device and the simplified one-dimensional model used for the simulations. (c)~Exemplary spatial distribution of $n_\mathrm{EL}$ and $n_\mathrm{WC}$ that is assumed for the simulations. }
    \label{fig:cycles}
\end{figure*}
Next we show that the entropy difference between the EL and WC phases can be harnessed to realise a continuous refrigeration cycle.
The heat extracted during an isothermal transition is given by ${Q}_\mathrm{cool} = \Delta S\,T$. 
Therefore, for $T_\mathrm{N} < T < T_\mathrm{SO}$ $\Delta S$ increases with decreasing $T$ [see Fig.~\ref{fig:pomeranchuk_nu4}(d)], which is favorable for cooling. 
Moreover, the EL--WC fraction is gate-tunable via the charge carrier density [Fig.~\ref{fig:pomeranchuk_nu4}(f)]; the transition can thus be driven electrostatically.
Building on these points, we design a continuous, gate-controlled cooling cycle.

The proposed cooling cycle is illustrated in the entropy–temperature ($S$–$T$) diagram shown in Fig.~\ref{fig:cycles}(a) and corresponds to an idealized isothermal process.
A possible experimental realization of this cycle is shown schematically in Fig.~\ref{fig:cycles}(b).
Flat-band graphene (MATBG or rMLG) is encapsulated in hexagonal boron nitride (hBN), providing an electrically insulating, atomically flat interface.
A bottom gate biases the device to a low-entropy phase with carrier density $n_\mathrm{EL}$ (assumed to be 100\% EL), while a local top gate drives a high-entropy state $n_\mathrm{WC}$ (100\% WC).
The local top gate allows for independent control of carrier density and displacement field in the WC region, enabling control over inversion symmetry breaking or the band gap~\cite{Andrei2020Dec,Banszerus2020May}, thereby avoiding unintended lifting of the flavor degeneracy.
An additional global top gate ensures that the EL regime can be tuned to high densities, ensuring that the WC melts and entropy is suppressed.
When a current flows through the device, the charge carriers undergo a heat cycle comprising the following steps:
\begin{itemize}[leftmargin=32pt]
    \item[(1$\rightarrow$2)] Carriers move from a low--entropy EL state ($n_\mathrm{EL}$) to a high--entropy WC state ($n_\mathrm{WC}$), absorbing heat from the electron system and lowering its temperature.
    \item[(2$\rightarrow$3)] Heat is advected along the device by the sliding WC beneath the top gate.
    \item[(3$\rightarrow$4)] Carriers return to a low--entropy EL state, releasing heat and locally raising the electron temperature.
    \item[(4$\rightarrow$1)] Carriers complete the loop and return to region (1), ready for the next cycle.
\end{itemize}
Because heat is absorbed on the left side of the top gate and released on the right side, a temperature gradient develops across the device.
An electronic payload bonded to the cold side is thereby cooled.
In the following subsections, we assess the device performance via simulations that compare the cooling power against Joule dissipation and parasitic heat leaks.

\subsection{Finite element model}
\begin{figure*}
    \centering   \includegraphics{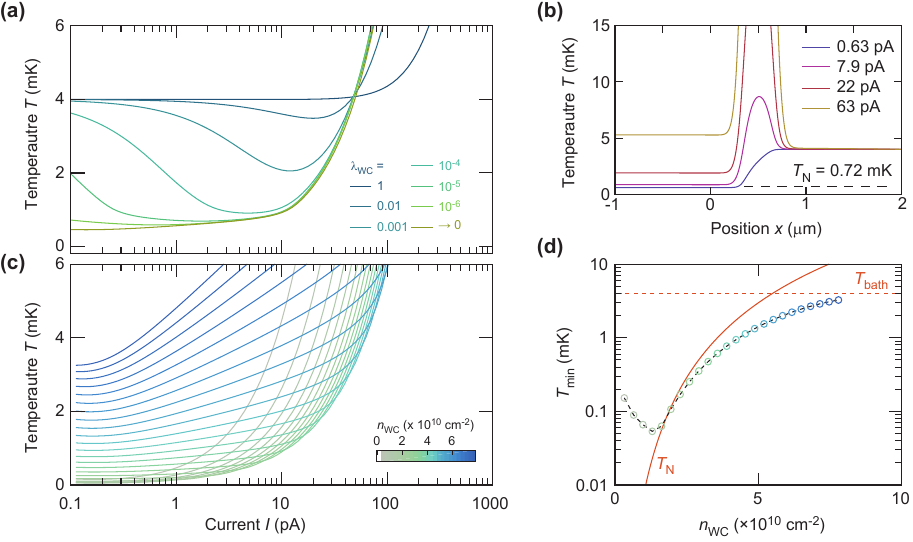}
        \caption{Simulations to determine the minimum possible electron temperature. Fixed simulation parameters are the electrical resistivity~$R_\mathrm{s} = 1~$\si{\kilo\ohm}, bath temperature~$T_\mathrm{bath} = 4$~\si{\milli\kelvin}.
        (a) Temperature in the cold region as a function of bias current for different values of the relative Lorenz number $\lambda_\mathrm{WC}$.
        (b) Exemplary temperature $T(x)$ for different values of the bias current $I$, for the case $n_\mathrm{EL} = \num{2.6e11}~$cm$^{-2}$, $n_\mathrm{WC} = \num{4.88e10}~$cm$^{-2}$ and $\lambda_\mathrm{WC} = \num{1e-6}$.
        (c)
        Temperature in the cold region as a function of bias current for different densities $n_\mathrm{WC}$ in units of cm$^{-2}$, assuming $\lambda_\mathrm{WC} = \num{1e-6}$. 
        (d) Minimal attainable temperature as a function of $n_\mathrm{WC}$. }
    \label{fig:bathleff}
\end{figure*}
To model the heat flow we assume a uniform current density across the device width, which justifies a one-dimensional finite-element model along the position $x$ such as illustrated in~Fig.~\ref{fig:cycles}(b). 
In calculating the drift velocity $v_\mathrm{d}(x)$, we assume fixed spatial distributions of the charge carrier densities $n_{\mathrm{EL}}(x)$ and $n_{\mathrm{WC}}(x)$ for the EL and WC phases, respectively, and impose current conservation throughout the device [see Fig.~\ref{fig:cycles}(c) and Appendix~B].
The temperature profile can then be obtained by solving the one-dimensional convection-diffusion equation that describes the transport of heat by charge carriers:
\begin{equation}\label{eq:convection_diffusion_main}
    -\partial_x (\kappa \partial_x T - v_\text{d} C_\mathrm{V} T) + \dot{q}_\text{V} = 0,
\end{equation}
where  $\kappa (x)$ is the thermal conductivity of the charge carriers, $C_\mathrm{V}(x)$ is the volumetric heat capacity and $\dot{q}_\mathrm{V}(x)$ is the volumetric heat flow. 
We model the local thermal conductivity and heat capacity by interpolating between their values in the pure EL and WC phases, weighted by the local phase fractions (see Appendix~B).
The volumetric source term is given by $\dot{q}_\mathrm{V} = \dot{q}_\mathrm{Joule}+\dot{q}_\mathrm{cool}+\dot{q}_\mathrm{phonon}$, corresponding to Joule heating, entropic cooling and electron-phonon coupling (see Appendix~B). 
We choose a geometry that is 3~\unit{\um} long, where the central 1-\unit{\um}-long section contains the WC, and the outer section contains the EL [see Fig.~\ref{fig:cycles}(b)]. 
The dimensions of the device were chosen to be representative of typical encapsulated graphene based devices reported in literature. 

A key performance factor is the WC thermal conductivity, which should be minimized to suppress heat backflow from the hot to the cold region.
In a diffusive electron gas, thermal and electrical transport are linked by the Wiedemann-Franz (WF) relation,
\begin{equation}
    \kappa = L_0 T \sigma,
\end{equation}
where $\sigma$ is the electrical conductivity and
$L_0 = \mbox{\qty{2.44e-8}{\volt\squared\per\kelvin\squared}}$ is the Lorenz number.
In the WC phase, however, electrical transport is governed by advection associated with the collective sliding of the crystal, whereas heat backflow from hot to cold is dominated by phonon conduction.
Because charge and heat transport proceed through distinct physical channels, the WF relation breaks down in the WC, leading to a strongly suppressed thermal conductivity, $\kappa_\mathrm{WC} \ll L_0 T \sigma_\mathrm{WC}$.
To introduce a physically intuitive scale for the WC thermal conductivity, we parametrize it as
\begin{equation}
\kappa_{\mathrm{WC}} = \lambda_\mathrm{WC} L_0 T \sigma_{\mathrm{WC}},
\end{equation}
where the dimensionless parameter $\lambda_\mathrm{WC}$ controls an effective Lorenz number $L_{\mathrm{eff}}^{\mathrm{WC}} = \lambda_\mathrm{WC} L_0$.
This expression is not meant to represent a microscopic WF law for the WC phase, but rather serves as a convenient phenomenological parametrization to explore the consequences of suppressed thermal transport relative to electrical transport.
For the EL phase, we assume that the standard WF relation remains valid.
In Appendix~C, we estimate that $\kappa_{\mathrm{WC}}$ is vanishingly small at the low temperatures considered here, primarily due to finite-size effects in our micron-scale device.
While the cooling mechanism itself does not rely on a specific device size, significantly larger structures may exhibit enhanced WC thermal conductivity and consequently reduced cooling (Appendix C).
In our simulations, we therefore vary $\lambda_\mathrm{WC}$ to assess how thermal backflow influences the device performance.

\subsection{Minimum attainable temperature}
To evaluate the minimum attainable temperature, we impose a zero-flux (Neumann) boundary at position $x = -1~\si{\micro\meter}$, and a fixed (Dirichlet) boundary $T = T_\mathrm{bath}$ at $x = 2~\si{\micro\meter}$ [Fig.~\ref{fig:bathleff}(a)].  
For a fixed current $I$, we solve Eq.~\eqref{eq:convection_diffusion_main} to obtain the temperature profile $T(x)$. 
Figure~\ref{fig:bathleff} shows results for a device sheet resistance $R_\mathrm{s} = 1$~\si{\kilo\ohm} and $T_\mathrm{bath} = 4$~\si{\milli\kelvin}, representative of state-of-the-art electron temperatures under dilution refrigeration~\cite{Bradley2016Jan}.
Figure~\ref{fig:bathleff}(b) shows example traces of $T(x)$ for various bias currents $I$. 
At very low currents (e.g. 0.63 pA), the device shows noticeable cooling, staying within 20\% of $T_\mathrm{N}$ up to a current of approximately 7.9~pA. 
Beyond this, the temperature rises as Joule heating becomes more prominent, with net heating observed for 63~pA. 
At this current, the maximum WC temperature rises to $61~\mathrm{mK}$ as a consequence of its low thermal conductivity. 
This is still well within WC region of the phase diagram [see Fig.~\ref{fig:pomeranchuk_nu4}(d)], so the WC should remain stable.
The transition from cooling at low currents to heating at higher currents is expected because the cooling contribution scales linearly with current, whereas Joule heating scales quadratically.

In Fig.~\ref{fig:bathleff}(a) we plot the average temperature over $-1$~\si{\micro\meter} $< x < 0$~\si{\micro\meter} as a function of $I$ for different $\lambda_\mathrm{WC}$. 
When $\lambda_\mathrm{WC} = 1$ (when the WC thermal conductivity follows the standard WF law), little cooling is observed.
This is because a high thermal conductivity drives backflow of heat from the hot region to the cold region, reducing the temperature gradient. 
Conversely, decreasing $\lambda_\mathrm{WC}$ suppresses backflow and yields progressively stronger cooling.
In the limit $\lambda_\mathrm{WC} \rightarrow 0$, the minimum temperature converges to a value close to the alignment temperature $T_\mathrm{N}$.

Figure~\ref{fig:bathleff}(c) shows $T(x=-1~\si{\micro\meter})$ as a function of $I$ for various values of $n_\mathrm{WC}$, with $\lambda_\mathrm{WC}$ fixed at $\num{1e-6}$.
The minimum temperature follows the trend set by $T_\mathrm{N}$ and $T_\mathrm{bath}$. 
However, two interesting deviations are noted [see Fig.~\ref{fig:bathleff}(d)]:
\begin{enumerate}
    \item Dense WCs ($n_\mathrm{WC} > \num{1.8e10}~$cm$^{-2}$) can exhibit an apparent undershoot, with $T_\mathrm{min}$ falling below both $T_\mathrm{N}$ and $T_\mathrm{bath}$. This occurs because $\mathrm{d}S/\mathrm{d}T$ peaks near $T_\mathrm{N}$ [Fig.~\ref{fig:pomeranchuk_nu4}(d)], increasing the heat capacity $C_\mathrm{V} = T~\mathrm{d}S/\mathrm{d}T$ and thereby enhancing the convection term in~Eq.~\eqref{eq:convection_diffusion_main}.  
    \item Dilute WCs ($n_\mathrm{WC} < \num{1.8e10}~$cm$^{-2}$) remain above $T_\mathrm{N}$ at finite $\lambda_\mathrm{WC}$ because the cooling power diminishes in very dilute crystals, making it harder to overcome the backflow and Joule heating. 
\end{enumerate}
The simulations indicate that sub-millikelvin temperatures are achievable only for sufficiently dilute WCs, in the example in Fig.~\ref{fig:bathleff}(d) for $n_\mathrm{WC} \lesssim \num{4.5e10}~$cm$^{-2}$ (corresponding to $r_s \gtrsim 42$).
However, extremely dilute WCs with densities close to $\num{0}~$cm$^{-2}$ show higher temperatures, which can be attributed to a reduction of cooling power. 
Therefore, the results seem to suggest a trade-off: higher density boosts cooling power, whereas lower density enables the lowest temperatures.
We explore the impact of carrier density on the cooling power in the next subsection.

\subsection{Cooling power of the device}
\begin{figure*}
    \centering   \includegraphics{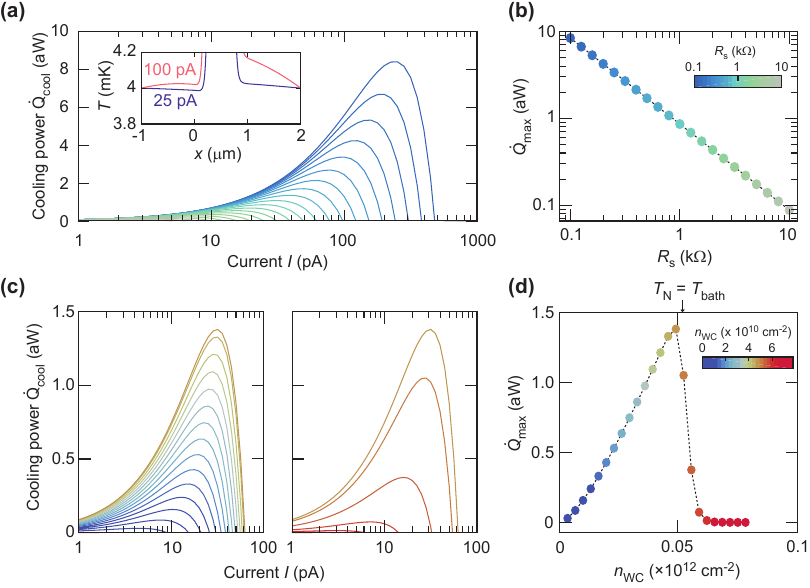}
    \caption{Cooling power and its dependence on sheet resistance, $T_\mathrm{bath} = 4~$mK and $\lambda_\mathrm{WC} = \num{1e-6}$. 
    (a) Cooling power $\dot{Q}_\mathrm{cool}$ versus current for several sheet resistances $R_\mathrm{s}$, the colours correspond to panel (b). Inset shows the temperature profile $T(x)$ for currents $I = 25~\si{\pico\ampere}$ and $I = 100~\si{\pico\ampere}$.
    (b) The maximum cooling power $\dot{Q}_\mathrm{max}$ as a function of $R_\mathrm{s}$. 
    (c) $\dot{Q}_\mathrm{cool}$ versus current for different densities $n_\mathrm{WC}$ (units cm$^{-2}$). With increasing $n_\mathrm{WC}$ the cooling power initially increases (left panel), reaches a maximum and then rapidly decreases (right panel). The colour scale corresponds to panel (d). 
    (d) Maximum cooling power $\dot{Q}_\mathrm{max}$ as a function of $n_\mathrm{WC}$. }
    \label{fig:coolingpower}
\end{figure*}
To estimate the device’s cooling power, we impose a Dirichlet boundary condition $T(x=-1~\si{\micro\meter}) = 4~$mK. 
The inset of Fig.~\ref{fig:coolingpower}(a) shows representative $T(x)$ traces. 
We obtain the cooling power from the total heat flux at the left boundary $\dot{Q}_\mathrm{cool} = (- \kappa \partial_x T)A_c$, where $A_c$ is the device cross-section. 
Figure~\ref{fig:coolingpower}(a) plots $\dot{Q}_\mathrm{cool}$ for several sheet resistances $R_\mathrm{s}$ as a function of the current $I$. 
The peak cooling power $\dot{Q}_\mathrm{max}$ depends strongly on $R_\mathrm{s}$, underscoring that lowering the sheet resistance is important for improving performance.
For $R_\mathrm{s}$ in the range \SIrange{0.1}{10}{\kilo\ohm}, we estimate a cooling power of \SIrange{0.08}{8.4}{\atto\watt} at $T_\mathrm{bath} = \SI{4}{\milli\kelvin}$ and optimal WC density. 
The effect of the $n_\mathrm{WC}$ is also evident in Fig.~\ref{fig:coolingpower}(c): $\dot{Q}_\mathrm{max}$ initially increases with $n_\mathrm{WC}$.
However, when $T_\mathrm{N}$ exceeds $T_\mathrm{bath}$ the magnetic moments in the WC become aligned, suppressing the entropy and causing $\dot{Q}_\mathrm{max}$ to drop sharply with increasing $n_\mathrm{WC}$ [inset of Fig.~\ref{fig:coolingpower}(d)].

For comparison of these cooling powers to the literature, we estimate the electron–phonon heat leak using the single-layer graphene model of Ref.~\cite{McKitterick2016Feb}, $\dot Q_{\mathrm{phonon}}=\Sigma A\,(T_{\mathrm e}^4-T_{\mathrm p}^4)$, where $\Sigma$ is a coupling constant.
Taking $\Sigma \approx 50~\mathrm{mW m^{-2} K^{-4}}$ at $T_{\mathrm e}=4~\mathrm{mK}$ and $T_{\mathrm p}=10~\mathrm{mK}$ over the entire device area, this gives $\dot Q_{\mathrm{phonon}}\approx 1.5~\mathrm{zW}$.
Based on this rough estimate, we argue that the most important heat load is likely from the metallic contacts. 
Using the WF relation, the minimum contact resistance $R_\mathrm{c}$ necessary to reduce the electron temperature $\Delta T$ by \SI{1}{\milli\kelvin} is given by: 
\[
R_\mathrm{c} = \frac{L_0\,T_\mathrm{bath}\,\Delta T}{\dot{Q}_\mathrm{cool}},
\]
implying $R_\mathrm{c}\sim \SIrange{7}{700}{\kilo\ohm}$ if $R_\mathrm{s}$ ranges between $\SIrange{0.1}{10}{\kilo\ohm}$.
Such large electrical contact resistances would be challenging in practice and reveal a central device-design constraint: the contacts must strongly suppress parasitic heat flow while still allowing low-dissipation current injection.
Therefore, contact materials with sub-Sommerfeld thermal conductivities, such as superconducting metals~\cite{Fong2013Oct} or correlated semimetals~\cite{Wang2025Jan}, could be employed to minimize parasitic heat leaks while keeping $R_\mathrm{c}$ low.

To enable fair comparison across on-chip coolers operating at different base temperatures, we use the entropy extraction rate $\Delta \dot{S}_\mathrm{ext} = \dot{Q}_\mathrm{cool}/T$.
For our device we find 0.022--2.1 fW/K, close to the experimentally reported value of $\sim1.8 - 5.4$ fW/K for a quantum dot refrigerator~\cite{Prance2009Apr}. 
Considering the considerably smaller volume of our atomically thin device, the efficacy of the cooling is competitive with the quantum dot refrigerator devices. 
The primary advantage of the WC-based cooler, however, is not the cooling power. 
Instead, it is the fact that the cooling remains effective in the sub-millikelvin regime, where other continuous coolers such as normal-insulating-superconductor (NIS) junctions and quantum-dot refrigerators lose efficiency.

\section{Discussion and conclusions}\label{sec:discussion}
In summary, we propose a cooling cycle that exploits the large entropy difference between the EL and WC phase to reduce electron temperatures to sub-millikelvin levels.
Under optimal conditions, our simulations indicate cooling powers between 0.08 and 8.4~aW per micrometre of device width, and electron temperatures well below 1 mK.
In addition, we highlight the effect of entropy on the EL–WC phase diagram, revealing ``freezing by heating'' reminiscent of the Pomeranchuk effect in $^3$He.

For the experimental implementation of this cooling cycle, several technical challenges must be addressed. 
High-quality samples with low electrostatic disorder are essential to minimise the resistance $R_\mathrm{s}$ caused by the sliding WC. 
To date, WC transport has been studied mainly in the pinned regime, where pinning opens a Coulomb gap and the resistance is very high~\cite{Shklovskii2004Jan}. 
With sufficiently low disorder and weak pinning, a WC driven above its depinning threshold may enter a collectively sliding regime with sufficiently low dissipation for the proposed cooling cycle~\cite{Zhu1994Aug,Brussarski2018Sep,Seiler2025Oct}.
Further research is necessary to determine which values of the sheet resistance $R_\mathrm{s}$ are realistic in this regime.
A second key experimental challenge is the thermal isolation from the metallic contacts that connect to the sample. 
Because the predicted cooling powers are in the attowatt regime, heat conduction from the contacts can easily overwhelm the cooling effect, which might be overcome by employing sub-Sommerfeld thermal conductivity materials.  
An optimal device geometry should additionally incorporate a short but wide top gate to reduce dissipation and enhance cooling power, while remaining sufficiently small to suppress heat backflow through phonon modes in the WC phase.
A wide device could also operate at larger total currents, relaxing the stringent demands on current-noise levels in the experimental setup.
Overall, realizing this cooling scheme will require simultaneous optimization of WC sliding transport, contact thermal isolation, and device geometry.

In principle, any flat-band, low-disorder material with a low-energy flavor degeneracy could host a WC phase suitable for sub-mK electron cooling.
In particular, multilayer graphene systems such as magic-angle twisted multilayers~\cite{Park2022Aug,Liu2022May} offer orbital and valley degeneracy and flat bands, while rhombohedral-stacked graphene~\cite{Holleis2024Jul,Han2023Nov} offer valley-degenerate flat bands under an applied displacement field.
Recently proposed moir\'e materials with low-energy states near the $M$-points---the midpoints of the edges of the hexagonal Brillouin zone---provide another promising platform, as their threefold valley degeneracy could yield an enhanced entropy in the WC phase~\cite{Calugaru2025Jul}.
However, an important consideration for multi-valley systems is that the exchange coupling may not be governed solely by pair- or ring exchange between neighboring carriers.
Recent work \cite{Calvera2023Jun} shows that effective-mass anisotropy can couple valleys through phonon zero-point motion in the WC, potentially leading to valley ordering and entropy reduction.
While the magnitude of this effect remains uncertain, it suggests that minimizing anisotropy --- for instance by careful material choice or small displacement fields --- could be important for maximizing the WC entropy.
For this reason, orbitally degenerate materials --- such as magic-angle twisted bilayer graphene near full filling of the flat bands --- may exhibit a more robust WC entropy when the combined twofold rotation and time-reversal symmetry ($\mathcal{C}_2\mathcal{T}$) is preserved \cite{Andrei2020Dec}.
Here, the degeneracy originates from states at the same crystal momentum (the $\Gamma_s$ point which is the center of the superlattice Brillouin zone), reducing sensitivity to effective-mass anisotropy and enhancing the robustness of the WC entropy.
Going forward, we suggest to investigate the temperature-dependent entropy of these materials combined with electronic transport experiments, to establish whether the proposed cooling cycle can, indeed, be established. 
Its realization in nanoelectronic devices could enhance quantum technologies and precision metrology, and open new avenues to explore correlated electron phenomena at ultra-low temperatures.

\acknowledgements
The authors thank N.~Drummond for sharing the data from Ref.~\cite{Drummond2009Mar} and thank K.~Ensslin, Y.M.~Blanter and S.~Rotkin for valuable discussions. This work was supported by the Deutsche Forschungsgemeinschaft (DFG, German Research Foundation) - 544940953, by the Deutsche Forschungsgemeinschaft (DFG, German Research Foundation) under Germany's Excellence Strategy – Cluster of Excellence Matter and Light for Quantum Computing (ML4Q) EXC 2004/2 – 390534769 and HFML-FELIX, a member of the European Magnetic Field Laboratory (EMFL).

\appendix

\section{Clausius-Clapeyron equation}
Considering two phases along the coexistence lines, we enforce the chemical potentials to be constant: $\mu_\mathrm{WC} = \mu_\mathrm{EL}$, where $\mu_\mathrm{WC},~\mu_\mathrm{EL}$ correspond to the WC and EL phase, respectively. 
 While maintaining constant pressure, we examine changes $\mathrm{d}n$ and $\mathrm{d}T$ along the coexistence curve, expressed as $\mu_i' = \mu_i + \mathrm{d}\mu_i$:
\begin{equation}
\begin{aligned}
\left( \frac{\partial \mu_\mathrm{WC}}{\partial n} \right)_{T} \mathrm{d}n  + \left( \frac{\partial \mu_\mathrm{WC}}{\partial T} \right)_{n} \mathrm{d}T  = \\
\left( \frac{\partial \mu_\mathrm{EL}}{\partial n} \right)_{T} \mathrm{d}n  + \left( \frac{\partial \mu_\mathrm{EL}}{\partial T} \right)_{n} \mathrm{d}T.
\end{aligned}
\end{equation}
Utilizing the Maxwell relation
\begin{equation}
    \left(\frac{\partial S_i}{\partial n}\right)_{T} = - \left(\frac{\partial \mu_i}{\partial T} \right)_{n}, 
\end{equation}
yields
\begin{equation}
\left[ \frac{\partial \mu_\mathrm{WC} }{\partial n} - \frac{\partial \mu_\mathrm{EL} }{\partial n}   \right]_T \frac{\mathrm{d}n}{\mathrm{d}T} = - \left[ \frac{\partial S_\mathrm{WC} }{\partial n} - \frac{\partial S_\mathrm{EL} }{\partial n} \right]_T.
\end{equation}
Integrating once over the carrier density leads to the Clausius-Clapeyron equation for the $n$--$T$ phase diagram:
\begin{equation}\label{eq:ClausiusClapeyronappendix}
\frac{\mathrm{d}n}{\mathrm{d}T} = - \frac{S_\mathrm{WC} - S_\mathrm{EL}}{\mu_\mathrm{WC} - \mu_\mathrm{EL}} ,
\end{equation}
where the integration constant has been omitted due to the limiting case where $S_\mathrm{WC} \equiv S_\mathrm{EL}$ resulting in ${\mathrm{d}n}/{\mathrm{d}T} =0$.

To determine $\mu_\mathrm{WC}$ and $\mu_\mathrm{EL}$, we use $\mu = \mathrm{d}E/\mathrm{d}n$, where $E$ is the internal energy. 
 These energies are obtained from the energy as a function of $r_\mathrm{s}$ using quantum Monte Carlo simulations from Ref.~\cite{Drummond2009Mar}. 
We assume that $\mu_\mathrm{EL}$ corresponds to the paramagnetic fluid and $\mu_\mathrm{WC}$ to the antiferromagnetic crystal as described within Ref.~\cite{Drummond2009Mar}. 

\section{Additional details on the finite element simulations}
\subsection{Entropy of the WC phase}
In our model, the entropy as a function of temperature for the solid Wigner phase is approximated by:
\begin{equation}
    \frac{S_\mathrm{WC}(T)}{k_B n} = \mathrm{ln}(2)\left[ \frac{1}{2} + \frac{1}{2} \tanh\left(\frac{4(T-T_\mathrm{N})}{T_\mathrm{N}}\right)\right].
\end{equation}
This function shows a step width corresponding to $0.5 k_B T$, consistent with Monte Carlo simulations \cite{Sauerwein1995Aug}. 
We note that this model is simplified to roughly capture $T_\mathrm{N}$ and to capture some key physical aspects, for example ensuring that
$S_\mathrm{WC} \approx 0$ and that $\mathrm{d}S_\mathrm{WC}/\mathrm{d}T \approx 0$ at $T = 0$.
Since our simulations focus on ultralow temperatures, effects of spin-orbit coupling and phononic contributions to the entropy are ignored. 

\subsection{Charge conservation}
While simulating we account for the two phases by defining a density $n_\mathrm{EL}(x)$ for the EL phase, and $n_\mathrm{WC}(x)$ for the WC phase, which both smoothly vary as a function of position. 
To take the stray fields of the top gate into account, we assume a smooth step with a width of $\gamma~=~50$ nm, giving the two density profiles:
\begin{equation}\label{eq:dens1}
n_\mathrm{EL}(x) = n_\mathrm{EL}^*-\frac{n_\mathrm{EL}^*}{2}\tanh{\left(\frac{x}{\gamma}\right)}+\frac{n_\mathrm{EL}^*}{2}\tanh{\left(\frac{x-L}{\gamma}\right)}
\end{equation}
\begin{equation}\label{eq:dens2}
    n_\mathrm{WC}(x) = n_\mathrm{WC}^*+\frac{n_\mathrm{WC}^*}{2}\tanh{\left(\frac{x}{\gamma}\right)}-\frac{n_\mathrm{WC}^*}{2}\tanh{\left(\frac{x-L}{\gamma}\right)}
\end{equation}
where $n_\mathrm{EL}^*$, $n_\mathrm{WC}^*$ are the uniform densities far away from the EL/WC interface. 
Note, that we choose to place the left side of the top gate at position $x = 0~\si{\micro\meter}$ and the right side is at position $x = L = 1~\si{\micro\meter}$.
An example of a density profile used in the simulation is shown in Fig.~\ref{fig:cycles}(c).
We varied the width of the transition region $\gamma$ in our benchmark simulations, and find that it has low impact on the temperature and was therefore not studied further.
The densities affect the drift velocity $v_\mathrm{d}$ as a function of position and since the current $I$ must be conserved, we can write:
\begin{equation}\label{eq:stokes}
    \frac{I}{W} = e v_\text{d}(x) n_\mathrm{EL}(x) + e v_\text{d}(x) n_\mathrm{WC}(x)
\end{equation}
where $W$ is the width of the sample.
Solving this equation for a fixed current $I$ directly yields the $v_\text{d}(x)$ term which is used in Eq.~\eqref{eq:convection_diffusion_main}.

\subsection{Thermal conductivity and heat capacity}
Our model also takes into account that both phases have different thermal conductivities $\kappa_1$ and $\kappa_2$, such that $\kappa(x)$ is a function of $x$. 
To accommodate this in the model, we describe the thermal conductivity as:
\begin{equation}
    \kappa(x) = k_1\frac{n_\mathrm{EL}(x)}{n_\mathrm{EL}^*}+ k_2\frac{n_\mathrm{WC}(x)}{n_\mathrm{WC}^*}.
\end{equation}
For the EL phase, we assume the thermal conductivity is determined by the WF relation, so $\lambda_\mathrm{EL} = 1$, while we vary $\lambda_\mathrm{WC}$ for the WC phase.

For the convection term, we need to take into account the volumetric heat capacity of both phases. 
For this, we use a similar approach as above and assume that:
\begin{equation}
    C_\text{V}(x) = \frac{\mathrm{d}S_\mathrm{EL}}{\mathrm{d}T} T\frac{n_\mathrm{EL}(x)}{n_\mathrm{EL}^*}+ \frac{\mathrm{d}S_\mathrm{WC}}{\mathrm{d}T} T \frac{n_\mathrm{WC}(x)}{n_\mathrm{WC}^*}.
\end{equation}
Note, that $\mathrm{d} S_\mathrm{WC}/\mathrm{d}T = 0$ when $T \gg T_N$, meaning that convection only impacts the heat transport through the WC phase near the alignment temperature $T_\mathrm{N}$.

\subsection{Heat fluxes}
Finally, we need to take into account the position-dependent heat flux $\dot{q}_\mathrm{V}$. 
We start with the heat source term due to the entropy change $\dot{q}_\mathrm{cool}$. 
If we consider an infinitesimal section $\mathrm{d}x$, we  find for the volumetric heat source $\dot{q}_\mathrm{cool}$:
\begin{equation}\label{eq:cooling}
    \dot{q}_\mathrm{cool} = - \frac{\Delta s\, T\, v_\text{d}}{h_\mathrm{g}}\,\frac{\mathrm{d} n_\mathrm{WC}}{\mathrm{d} x},
\end{equation}
where $h_\mathrm{g} = 0.67~$nm is the thickness of the MATBG layers. 
Note that, when ${\mathrm{d} n_\mathrm{WC}}/{\mathrm{d} x} > 0$, $\dot{q}_\mathrm{cool}<0$ results in net cooling power while if ${\mathrm{d} n_\mathrm{WC}}/{\mathrm{d} x} < 0$,  $\dot{q}_\mathrm{cool}>0$ results in the release of heat. 
In addition, we take the effect of Joule heating into account
\begin{equation}
    \dot{q}_\text{Joule} = \frac{R_s I^2}{W^2 h_g}
\end{equation}
where $h_g \approx 0.67$~nm is the thickness of the MATBG. 
For simplicity, we take the square resistivity $R_s$ to be uniform over the sample geometry, regardless of the phase. 
Finally, we need to take the heat flow between the electron and phonon bath into account. 
To do this, we use the results obtained in Ref.~\cite{McKitterick2016Feb} on single-layer graphene, and assume $\dot{q}_\text{phonon} = \Sigma (T_\text{e}^4 - T_\text{p}^4)$, where we take $\Sigma = 50$ mW/m$^2$K$^4$. 
We assume this to be uniform over the sample geometry for both phases. 

\section{Ballistic phonon heat transport in a Wigner crystal}
To estimate the backflow of heat mediated by phonons in the WC phase, we consider a simplified ballistic model, which provides an upper bound on the thermal conductivity by neglecting scattering.
The phonon dispersion of a WC contains 2 branches~\cite{Bonsall1977Feb}, which can be described by:
\begin{equation}\label{eq:omega1}
    \omega_1(k) = \omega_p \sqrt{a_0 k} [1- C (a_0k)]
\end{equation}
\begin{equation}
    \omega_2(k) = B \omega_p a_0 k 
\end{equation}
where $a_0$ is the length of the primitive Bravais lattice vector, $k$ is the momentum, $B = 0.1905$, $C = 0.09074$ are numerical constants and $\omega_p$ a characteristic frequency given by:
\begin{equation}
    \omega_p^2 = \frac{e^2}{2 \varepsilon_0 \varepsilon_r' m^* A_c a_0 }
\end{equation}
where $A_c$ is the primitive unit cell area (of the Wigner crystal). Expressing this in terms of the charge carrier density we obtain:
\begin{equation}
    \omega_p = \frac{\sqrt[4]{3} e^2 n^{3/2}}{2 \sqrt{2} \varepsilon_0 \varepsilon_r' m^*}.
\end{equation}

\begin{figure*}
    \includegraphics{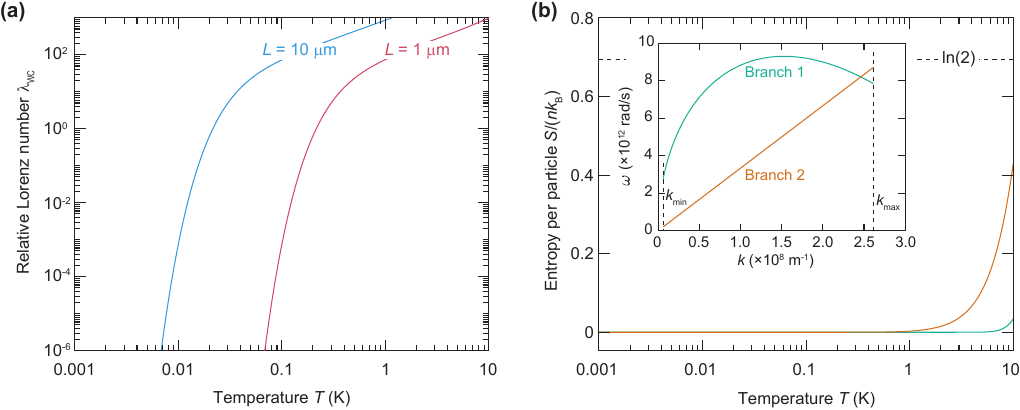}
    \caption{(a) Relative Lorenz number $\lambda_\mathrm{WC}$ versus temperature for a sample lengths $L=\SI{1}{\micro\meter}$ and $L=\SI{10}{\micro\meter}$. (b) Contribution of the phonon branches to the entropy per particle. Inset shows the two phonon branches.}
    \label{fig:appendix_conductivity}
\end{figure*}
Using the dispersion relations, we can calculate the internal energy of each branch as:
\begin{equation}\label{eq:internalenergy}
    U_j =\int\limits_{k_\mathrm{min}}^{k_\mathrm{max}} \frac{k}{2\pi}  \frac{\hbar \omega}{\mathrm{e}^{\hbar \omega/k_B T}-1} \mathrm{d} k
\end{equation}
The integration limits are determined by either the sample size $L$: $k_\mathrm{min} = {2\pi}/{L}$ or the lattice constant of the WC $a_0$: $k_\mathrm{max} = {2\pi}/{a_0}$.
To estimate the thermal conductivity in the ballistic limit, we assume a heat flux $Q_j = U_j v_j = U_j \mathrm{d}\omega/\mathrm{d}k$ is emitted by each phonon branch:
\begin{equation}
    Q_j = \int\limits_{\omega_\mathrm{min}}^{\omega_\mathrm{max}} \frac{k}{2\pi h_g}  \frac{\hbar \omega}{\mathrm{e}^{\hbar \omega/k_B T}-1} \mathrm{d} \omega.
\end{equation}
Where $\omega_\mathrm{min}$ and $\omega_\mathrm{max}$ are determined from the dispersion relations. Next, this function is evaluated at different temperatures $T_1$ and $T_2$, and the effective thermal conductivity follows from Fourier's law:
\begin{equation}
    \kappa_{\mathrm{eff},j} = \frac{Q_j(T_2)-Q_j(T_1)}{\left( \frac{T_2-T_1}{L} \right)}
\end{equation}
and the total thermal conductivity $\kappa_\mathrm{eff} =\kappa_{\mathrm{eff},1} + \kappa_{\mathrm{eff},2}$.
Fig.~\ref{fig:appendix_conductivity}(a) shows $\kappa_\mathrm{eff}$ as a function of temperature for a sample size $L = 1~$\si{\micro\meter} and $L = 10~$\si{\micro\meter}. 
For the micron sizes example used in the main text, the thermal conductivity is vanishingly small at temperatures below $\sim$ 70 mK.
However, increasing the sample size considerably increases the thermal conductivity.

\subsection{Phononic contribution to WC entropy}
From Eq.~\ref{eq:internalenergy} we can also evaluate the volumetric heat capacities of each phonon branch as:
\begin{equation}
    C_{V,j} = \frac{\mathrm{d}U_j}{\mathrm{d}T}
\end{equation}
Next we can integrate the relation (at constant volume and particle number):
\begin{equation}
    C_{V,j} \mathrm{d}T = T \mathrm{d} S_j
\end{equation}
and assume $S_j|_{T=0} = 0$ to obtain the phononic contributions to the entropy shown in Fig.~\ref{fig:appendix_conductivity}(b).

\bibliography{Pomeranchuk_literature}

\end{document}